\documentclass[twocolumn]{revtex4}[12pt]
\usepackage{graphicx,multirow,array,amsmath,empheq,amsthm, amssymb,amsfonts,tikz,ragged2e,color}
\usepackage{graphicx}
\newcommand{\beq}{\begin{equation}}
\newcommand{\eeq}{\end{equation}}
\newcommand{\beqa}{\begin{eqnarray}}
\newcommand{\eeqa}{\end{eqnarray}}
\begin{document}
\title{Electron-Acoustic Solitons in Magnetized Collisional Nonthermal Plasmas}
\author{M. R. Hassan$^{1}$, T. I. Rajib$^{1}$\footnote{Email: tirajibphys@juniv.edu OR tirajibphys@gmail.com}, S. Sultana$^{1}$}
\affiliation{$^1$Department of Physics, Jahangirnagar University, Savar, Dhaka-1342, Bangladesh}
\begin{abstract}
 The properties of obliquely propagating dissipative electron-acoustic solitary waves (OPdEASWs) have been investigated in a magnetized collisional superthermal plasma consisting of inertial cold electrons, inertialess hot electrons featuring $\kappa-$distribution and static ions via the fluid dynamical approach. By using the reductive perturbation technique, a nonlinear Schamel equation that governs the nonlinear features of OPdEASWS is obtained. The solitary wave solution of the Schamel equation is used to examine the basic features of small, but finite amplitude OPdEASWs in such a magnetized collisional superthermal plasma. The basic features (width, amplitude, speed, etc.) of OPdEASWs are found to be modified by the different plasma configuration parameters, such as the superthermality index, obliquity angle, collisional parameter, trapping parameter, and external magnetic field.  The nature of electrostatic disturbances, that may propagate in different realistic space and laboratory plasma systems (e.g., in Saturn ring), are briefly discussed.
\end{abstract}
\maketitle

\section{Introduction}
\label{introduction}
Plasma has different inherent characteristics such as dispersion,
nonlinearity, and dissipation. The localized structures of permanent profile (or solitary waves) are considered
as a consequence of the balance between the nonlinear
and the dispersive effects with weak dissipation. These nonlinear structures (or solitons) experience dissipation while propagating in the dissipative plasma medium by changing their amplitude, width, speed, and eventually diminish with time. The nonlinear propagation of dissipative solitary waves (or dissipative solitons) have received a great deal of attention in plasma physics \cite{Nicholson1976,Pereira1978,Dutta2012,Ghosh2014,SS2015} and also in nonlinear optics \cite{Mollenauer1980,Nozaki1983,Afanasjev1995,Malomed1996,Atai1996,Soto1997}, because of the real physical systems in nature are far away from equilibrium, and exhibit dissipative properties. In plasmas, the dissipation may arise due to the collisions of various plasma constituents (e.g., the electron-neutral collision \cite{Krapak2001}, the ion-neutral collision \cite{Vladimirov1999}, the dust-ion collision \cite{Vladimirov1999}, the dust-neutral collision \cite{Krapak2001}, etc.) or due to nonlinear Landau damping \cite{Chang1970} or due to kinematic fluid viscosity \cite{Nosenko2004}. It can be mentioned that the solitons are used for optical fiber communications \cite {Haus1996} because of their excellent nature of stationary profile. In a dissipative system, an external source of energy (amplifier, amplitude modulator) \cite{Nakazawa1991} is needed to send the original information over  very long distances in fiber optic communications. The external energy balances the dissipation, and the initial solitary pulse maintains it's stability while propagating.

Electron-acoustic waves (EAWs) \cite{Watanabe1977,Yu1983,Toker1984,Gary1985,Mace1990,IK2004,AAMamun2002,Mace1999,Baluku2010,SS2011,SSI2012,SSIK2012}, which may propagate in a plasma containing cold and hot electrons, where the temperature of hot electrons is a minimum ten times higher than that of cold electrons \cite{Toker1984,Mace1999,SS2011}, are seen both in space and laboratory plasma environments \cite{Montgomery2001,Lu2005,Dieckmann2007,Rothkaehl2009}. EAWs are high frequency electrostatic modes, where inertia (restoring force) is provided by the mass density (thermal pressure) of cold electrons (hot electrons) and ions only maintain the background charge neutrality of the plasma. This kind of high-frequency wave's phase speed lies in the range between the thermal speed of the cold and hot electron species to avoid the resonant damping \cite{Toker1984,Mace1999,SS2011}. It is noted here that the Landau damping will be minimized if the plasma the medium consists of cold electron populations approximately from a fifth to four-fifths of the total number of electrons, and to exhibit the propagation of EAWs is possible in that plasma. The propagation of EAWs has been reported in the middle-altitude cusp region \cite{Rothkaehl2009} by instruments onboard CLUSTER spacecraft and also in the laboratory experiment \cite{Montgomery2001}. The computer experiment (numerical simulation) has also confirmed the propagation of EAWs via the particle-in-cell simulations as a result of beam-driven instabilities \cite{Lu2005,Dieckmann2007}. It has also seen that EAWs can be driven by electron beams in a plasma comprising a hot and a cold electron population \cite{Lu2005,Ali2019}. Therefore, the propagation nature of EAWs seems to be a growing topic of research interest because of the existence of EAWs in space and laboratory plasma environments, and the linear and nonlinear properties of EAWs have already been investigated by many authors in different plasma contexts \cite{Watanabe1977,Yu1983,Toker1984,Gary1985,Mace1990,IK2004,AAMamun2002,Mace1999,Baluku2010,SS2011,SSI2012,SSIK2012,Montgomery2001,Lu2005,Dieckmann2007,Rothkaehl2009,Danehkar2011,Singh2004,Matsumoto1994,Ergun1998,Pickett1999,SS2015,SS2018,mamun2002,m2012,Bansal2019a}.

Electron trapping is now an observed phenomenon both in space and laboratory plasmas \cite{Ergun1998,ander02,cat03,S2019,schip08,ergun02}, where some of the plasma particles are confined and bounce back and forth in a finite region of space. It is noted that electrons do not follow the thermal Maxwellian distribution due to the formation of phase space holes caused by the trapping of electrons in the wave potential. Particle trapping has already been confirmed analytically, numerically, space plasma observation, and laboratory experiments \cite{Ergun1998,ander02,cat03,schip08,ergun02,lynov79,god03}. The electron distribution in the presence of trapped particles was introduced via the Schamel vortex-like electron distribution \cite{scham72,scham73} to model the KdV-like Schamel equation \cite{scham73,verheest94,Williams2014} for weakly nonlinear ion-acoustic waves. After that, a large number of theoretical works \cite{AAMamun2002,mamun2002,mamun96,nejoh97,mamun98,Tribeche2012,hafez16} has been carried out to study the nonlinear wave (solitary waves, shocks, etc.) properties in different plasma contexts in the presence of trapped particles. Recently, Williams {\it{et al.}} \cite{Williams2014}, modelled and characterized the ion-acoustic solitary waves via the Schamel equation \cite{verheest94} in an unmagnetized electron-ion plasma, while Sultana {\it{et al.}} in Ref. \cite{sultana19} studied the properties of obliquely propagating ion-acoustic solitary waves in a magnetized plasma with $\kappa$-nonthermal trapped electron populations. Employing reductive perturbation theory Korteweg-de-Vries-Zakharov-Kuznetsov evolution equation is derived to investigate obliquely propagating EASWs in a plasma holding nonthermal hot electrons, cold and beam electrons, and ions in a magnetized plasma \cite{singh2016}. The impact of electron trapping and nonextensivity on EASWs have been discussed in Ref. \cite{Shan2016}. Zakharov-Kuznetsov equation for ion-acoustic waves is derived using the reductive perturbation method in magnetised plasmas with nonthermal electrons and positrons \cite{Chawla2020}. Theoretical and numerical investigations are studied for the oblique modulation of EAWs in three-component plasma by Bansal {\it{et al.}} \cite{Bansal2019}. Sultana \cite{S2019} examined the obliquely propagating EASWs in collision-free magnetized plasma with trapped superthermal electrons as well. Shan \cite{Ali2019} investigated the dissipative electron-acoustic solitons in a cold electron beam collisional unmagnetized plasma with superthermal trapped electrons. Mouloud \cite{m2012} worked with the EASWs in a magnetized plasma with hot electrons featuring Tsallis distribution and his findings are relevant to the dayside auroral zone. The influence of collisions on the dynamics of a cold non-relativistic plasma has been studied by Brodin and Stenflo \cite{Brodin2017}. So, we can consider both the effect of the magnetic field and collision in case of EASWs propagation. We are interested in analyzing the evolution characteristics of OPdEASWs via the Schamel equation in a magnetized collisional plasma with $\kappa-$distributed trapped superthermal electrons. So, relying on our plasma fluid model, we model the oblique propagation of dissipative EASWs (dEASWs) via the Schamel equation in a magnetized $\kappa$-nonthermal plasma, and analyze the basic features and evolution characteristics of OPdEASWs theoretically and numerically. The main focus here is to investigate the influence of particle trapping on the dynamics of OPdEASWs, and also to analyze the effect of ambient magnetic field and the role of damping parameter on the characteristics of obliquely propagating electron-acoustic solitary excitations.

The layout of this article is as follows: The theoretical plasma model for electrostatic electron-acoustic waves is presented in section \ref{Model}. The Schamel equation for OPdEASWs is derived by adopting the reductive perturbation method in section \ref{KdV-Schamel}. The solitary wave solution is obtained via the hyperbolic tangent approach and nonlinear analysis, depending on the parametric and numerical investigations, is carried out for small but finite amplitude OPdEASWs in the subsections \ref{easoln} and \ref{parainvest}.  Finally, our theoretical and numerical results are briefly discussed in section \ref{Discussion}.

\section{Theoretical Plasma Model} \label{Model}
The magnetized collisional plasma system consisting of inertial cold electron fluid, noninertial $\kappa$-distributed hot electrons, and stationary ions is considered. Thus, the charge neutrality condition for our plasma model can be written as  $Z_in_{i0}=n_{c0}+n_{h0}$, where $n_{i0}$, $n_{c0}$, and $n_{h0}$ are the unperturbed number density of the ions, cold electrons, and hot electrons, respectively. $Z_i$ is the number of protons residing onto the ion surface. The ambient magnetic field $\mathbf{B}_0$ is considered  to lie along the $z$-axis, i.e., $\mathbf{B}_0=B_0\mathbf{\hat{z}}$ (here, $\mathbf{\hat{z}}$ is a unit vector along the $z$-axis). Thus, in such a magnetized collisional plasma medium, the dynamics of OPdEAWs, whose phase speed is much greater than both ions' and cold electrons' thermal speed, and far smaller than hot electrons' thermal speed, i.e., $v_{th,h}\gg v_{ph}\gg v_{th,i,c}$, can be described by the following fluid dynamical equations 
\begin{eqnarray}
&&\hspace*{-0.5cm}\partial_t n_{c}+\mathbf{\nabla}.(n_{c}\mathbf{u}_{c})=0,
\label{eqd1}\\
&&\hspace*{-0.5cm}\partial_t \mathbf{u}_{c}+(\mathbf{u}_{c}.\nabla)\mathbf{u}_{c}=\frac{e}{m_e}\mathbf{\nabla}\phi-\frac{e B_0}{m_e}\left(\mathbf{u}_{c}\times\mathbf{\hat{z}}\right)-\nu_{en}{\bf u}_c,
\label{eqd2}\\
&&\hspace*{-0.5cm}\nabla^{2}\phi=4\pi e \left[n_c +n_h-Z_i n_{i0}\right],
\label{eqd3}\
\end{eqnarray}
where $n_c$ ($n_h$) is the number density of cold electrons (hot electrons).  ${\bf u}_c$, $\phi$, and $\nu_{en}$ are, respectively,  the cold electron fluid velocity, electrostatic wave potential, and the cold electron-neutral collision frequency. $m_e$ ($e$) is the mass (magnitude of the charge) of an electron.

In order to separate the free electrons from the trapped, we introduce the concept of energy separatrix \cite{scham72,Tribeche2012,Schmel1975}. The energy separatrix occurs at the point where the energy of the electrons  $E_e$ equals to zero. The free electrons have $E_e>0$ and for the trapped electrons $E_e<0$. The presence of trapped electron population is considered via the $\kappa-$type non-Maxwellian (superthermal) the distribution function for the hot electrons \cite{Williams2014}
\begin{eqnarray}
&&\hspace*{-0.5cm}f^{\kappa}_{t}(v)=\frac{\Gamma(\kappa)}{\sqrt{2\pi}(\kappa-3/2)^{1/2}\,\Gamma(\kappa-1/2)}\nonumber\\
&&\hspace*{0.5cm}\times\left[1+\alpha\left(\frac{v^2/2-\Phi}{\kappa-3/2}\right)\right]^{-\kappa},
\label{tkpdis}\
\end{eqnarray}

where $\alpha=T_{hf}/T_{ht}$ is a parameter (ratio of the thermal energy of the free hot electrons to the thermal energy of the trapped hot electrons), which determines the efficiency of electron trapping \cite{Shan2018}, $\kappa$ is the superthermal index which measures the superthermality of the particles in the distribution function, and $\Phi=e\phi/T_h$ (in which $T_h$ is the hot electron thermal energy). It is essential to mention here that the Vasyliunas-Schamel distribution function is not valid for a magnetized plasma in general. However, it is valid due to the fact that the strong magnetic field and a small mass of electrons, their Larmor radii are so small as if they flow along the magnetic lines of force and do not feel the effect of the external magnetic field. The number density of hot electrons following the $\kappa$-nonthermal trapped distribution function $f_{t\kappa}= f^{\kappa}_{t}(v,\phi)+ f^{\kappa}_f(v,\phi)$ [where $f^{\kappa}_f(v,\phi)$ is the $\kappa$-distribution function for free electrons presented in details by Hellberg {\it et al.} \cite{hell09}, and is  valid for two ranges of the electron speed, namely ($-\infty$, $-\sqrt{2\phi}$) and
($\sqrt{2\phi}$, $\infty$)] is given by \cite{Williams2014}

\begin{eqnarray}
&&\hspace*{-0.5cm}n_h(\phi)=\int^{-\sqrt{2\phi}}_{-\infty}\,f^{\kappa}_f(v,\phi)dv+\int^{\sqrt{2\phi}}_{-\sqrt{2\phi}}\,f^{\kappa}_{t}(v,\phi)dv\nonumber\\
&&\hspace*{0.6cm}+\int^{\infty}_{\sqrt{2\phi}}\,f^{\kappa}_f(v,\phi)dv.
\label{tkn}\
\end{eqnarray}

Now, the dynamics of such a plasma system can be described by the set of normalized continuity, momentum, and Poisson's equations in the following forms
\begin{eqnarray}
&&\hspace*{-0.5cm}\partial_T n+\mathbf{{\tilde \nabla}}\cdot(n\mathbf{u})=0,
\label{eq1}\\
&&\hspace*{-0.5cm}\partial_T\mathbf{u}+(\mathbf{u}\cdot {\tilde\nabla})\mathbf{u}=\mathbf{{\tilde\nabla}}\Phi -\Omega_c \left(\mathbf{u}\times\mathbf{\hat{z}}\right)-\nu_e{\bf u},
\label{eq2}\\
&&\hspace*{-0.5cm}{ \tilde\nabla}^{2}\Phi\simeq \beta(n-1)+N_h-1,
\label{eq3}\
\end{eqnarray}
where the normalizing and associated parameters are defined as: $N_h=n_h/n_{h0}$,  $n=n_c/n_{c0}$, ${\bf u}={\bf u}_c/c_0$ with characteristic sound speed $c_0=\sqrt{T_h/m_e}$, $\Phi=e\phi/T_h$, $\mathbf{{\tilde \nabla}}=\lambda_{Dm}\mathbf{\nabla}$ [in which characteristic Debye length
$\lambda_{Dm}=\sqrt{(T_h/4\pi e^2 n_{h0})}$], $T=t\omega_{ph}$ in which plasma frequency $\omega_{ph}=\sqrt{(4\pi e^2  n_{h0}/m_e)}$, $\Omega_c=\omega_{ch}/\omega_{ph}$ with cyclotron frequency $\omega_{ch}=e B_0/m_e$, $\nu_e=\nu_{en}/\omega_{ph}$, and cold-to-hot electron number density ratio at equilibrium $\beta=n_{c0}/n_{h0}$.

The normalized number density expression of hot electron species $N_h$ which can be obtained by  performing all the integrations in (\ref{tkn}) appropriately has the form
\begin{equation}
N_h=1+\mathcal{S}_1 \Phi+ \mathcal{S}_2 \Phi^{3/2}+\mathcal{S}_3\Phi^2+\cdots,\label{eden}
\end{equation}
where $\mathcal{S}_1$, $\mathcal{S}_2$, and $\mathcal{S}_3$ are given by
\begin{eqnarray}
&& \mathcal{S}_1=\frac{2\kappa-1}{2\kappa-3},\qquad \mathcal{S}_2=\frac{8\sqrt{2/\pi}(\alpha-1)\,\kappa\,\,\Gamma(\kappa)}{3(2\kappa-3)^{3/2}\,\Gamma(\kappa-1/2)},\nonumber\\
&&\mathcal{S}_3=\frac{4\kappa^2-1}{2(2\kappa-3)^2}.\nonumber
\end{eqnarray}

We note that the number density for $\kappa$-distributed superthermal hot-electron species (i.e., $N_h\simeq 1+\mathcal{S}_1 \Phi+\mathcal{S}_3\Phi^2$)\cite{SSI2012,sarri10} is recovered for the limiting case $\alpha\rightarrow 1$, and the number density for the vortex-like Schamel distributed Maxwellian electrons (i.e., $N_h\simeq1+\Phi+ \frac{4(\alpha-1)}{3\sqrt{\pi}} \Phi^{3/2}+\frac{\Phi^2}{2}+\cdots$) \cite{scham73} is recovered for the limiting case $\kappa\rightarrow\infty$. That is, the plasma model under consideration in this manuscript is a generalized model to analyze the effect of $\kappa$ superthermal trapped electrons, trapped Maxwellian electrons on the dynamical properties of OPdEASWs for different limiting cases.

\section{\lowercase{d}EASW\lowercase{s} \lowercase{in} magnetized plasmas} \label{KdV-Schamel}
To study the existence of finite-amplitude OPdEASWs in a magnetized collisional plasma under consideration, the independent coordinates are stretched as
\cite{Williams2014,ferdousi15}
\begin{equation}
\xi=\epsilon^{1/4}\,\left(l_x x+l_y y+l_z z-v_p t\right)\,,\,\,\,\,\tau=\epsilon^{3/4}\,T\,,\label{cor}
\end{equation}
where $\epsilon$ is a negligibly small parameter ($0<\epsilon<1$)  which measures the strength of the nonlinearity, $v_p$ is the electron-acoustic wave phase speed normalized by $c_0$, and $l_x,\,l_y$, and $l_z$ are the directional cosines of the wave vector $\mathbf{k}$ having $x,\,y,$ and $z$-components, respectively (i.e., $l_x^2+l_y^2+l_z^2=1$). We should note that $x,\,y,\,z$ are all normalized by $\lambda_{Dm}$ and $\tau$ is normalized by the hot electrons' plasma period $\omega_{ph}^{-1}$. To model the dEASWs, we consider the presence of small damping in the plasma because of electron-neutral collision by assuming $\nu_e=\epsilon^{3/2}\nu$. All the dependent physical quantities $n,\,\mathbf{u}$, and $\Phi$ can be expressed around their equilibrium values in terms of $\epsilon$ as follows
\begin{eqnarray} \left.
\begin{array}{l}
n=1+\epsilon n_{1}+\epsilon^{3/2} n_{2}+ \cdots \ , \\
u_{x,y}=\epsilon^{5/4} \,u_{1x,y}+\epsilon^{3/2}\, u_{2x,y}+ \cdots \ , \\
u_{z}=\epsilon \,u_{1z}+\epsilon^{3/2}\, u_{2z}+ \cdots \ , \\
\Phi=\epsilon \phi_{1}+\epsilon^{3/2} \phi_{2} + \cdots .
\end{array}
\right\} \label{expan}
\end{eqnarray}
We now substitute  (\ref{eden}), (\ref{cor}),  and (\ref{expan}) into (\ref{eq1})-(\ref{eq3}). The expression for the phase speed (which leads us to analyze the linear EAW properties) is obtained by separating the coefficient of the lowest order of
$\epsilon$ from the resultant equations (i.e., $\epsilon^{5/4}$ from the continuity equation and the $z$-component of momentum equation, and $\epsilon$ from Poisson's equation) as
\begin{equation}
v_p =| l_z|\sqrt {\beta\left({\frac{2\kappa-3}{2\kappa-1}}\right)} = |l_z|\sqrt{\beta/\mathcal{S}_1}\, . \label{phase}
\end{equation}

\begin{figure}[!h]
\centering
\includegraphics[width=8.25cm]{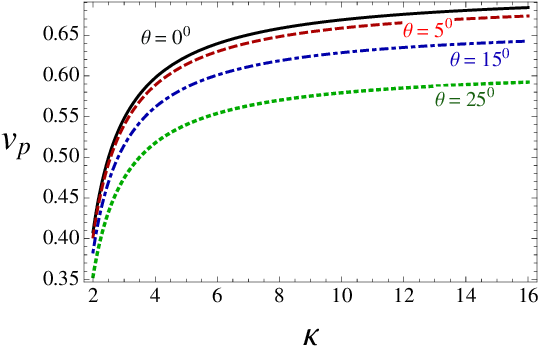}

\large{ (a)}
\vspace{0.5cm}

\includegraphics[width=8.25cm]{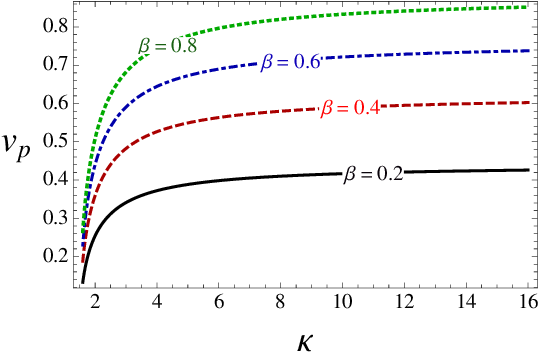}

\large{ (b)}
\caption{Showing the  variation of the phase speed $v_p$  versus the superthermality index $\kappa$ for (a) different obliquity angle $\theta$, where cold-to-hot electron number density at equilibrium $\beta=0.5$, and (b) for different $\beta$, where $\theta=10^{\circ}$.} \label{Fig1}
\end{figure}

It is clear from the equation (\ref{phase}) that the phase speed $v_p$ of EAWs in a magnetized plasma depends on the obliquity angle $\theta$ via $\theta~(=cos^{-1}|l_z|)$, plasma superthermality (nonthermality) via the superthermality index $\kappa$, and also on the cold-to-hot electron number density at equilibrium $\beta$. The variation of $v_p$ with  $\kappa$ for different values $\theta$ is depicted in Figure \ref{Fig1}(a), while the variation of $v_p$ with $\kappa$ for different values of $\beta$ is shown in Figure \ref{Fig1}(b). It is seen in both panel of Figure \ref{Fig1} that the phase speed $v_p$ is lower in a strongly superthermal plasma (lower $\kappa$) and the phase speed is higher in a Maxwellian plasma (higher $\kappa$), which agrees with earlier investigations for different plasma contexts \cite{SS2011,SSIK2012,sultana19,sultana10}. It is also found that the phase speed of EAWs in a magnetized superthermal plasma decreases with the obliquity angle $\theta$, i.e., the phase speed $v_p$ is higher for parallel propagating EAWs whereas lower for obliquely propagating EAWs, as displayed in Figure \ref{Fig1}(a). However, the effect of inertial cold electron number density (via $\beta$) on $v_p$ is illustrated in Figure \ref{Fig1}(b), in which we can see that the increase in the inertial cold electron number density in the plasma medium increases the phase speed of EAWs in the considered plasma. It is also essential to mention that the positivity of $v_p$ indicates the linear stability of EAWs for $\kappa>3/2$, $0<\theta<90^\circ$, and also for $0.25\leq\beta\leq 4$. We note that we consider the ranges of density ratio $0.25\leq\beta\leq 4$ \cite{Toker1984,SS2011,mace99} for which the phase of EAWs lies in between the thermal speed of hot and cold electrons to avoid the resonant damping, i.e.,  the propagation of EAWs may not possible in a plasma medium with cold electron population less than a quarter that of hot electron population.

The $x$ and $y$-components of the electric field drift of the electron fluid, in terms of the electric potential $\phi_1$, can be obtained by separating the $y$ and $x$-components of the momentum equation, respectively, are given by
\begin{eqnarray}
&&\hspace*{1.0cm}u_{1x,y}=\mathcal{R}_{1x,y}\partial_{\xi} \phi_1,\label{u1x}
\end{eqnarray}
where $\mathcal{R}_{1x,y}=\mp\frac{l_{y,x}}{\Omega_c}$ and $-(+)$ represents the $x$ and $y$-components of the electric field drift of the electron fluid. The next higher order of $\epsilon$ (i.e., separating $\epsilon^{3/2}$ from momentum equation) leading to the $x$ and $y$-components of the polarization drift of the electron fluid in the form
\begin{eqnarray}
&&\hspace*{1.0cm}u_{2x,y}=\mathcal{R}_{2x,y}\partial_{\xi}^2 \phi_1,\label{u2x}
\end{eqnarray}
where $\mathcal{R}_{2x,y}=\mp\frac{v_p}{\Omega_c}\mathcal{R}_{1y,x}$.
Following the same procedures (i.e., separate the coefficients of $\epsilon^{7/4}$ from the continuity and  $z$-component of the momentum equation, and $\epsilon^{3/2}$ from  Poisson's equation) and eliminating $n_2,\,u_{2z}$, and $\phi_2$ from the resultant equation, one can obtain a nonlinear partial differential equation in the form of Schamel equation as
\begin{equation}
 {\partial_\tau}\psi+A\psi^{1/2}{\partial_\xi}\psi+B{\partial_{\xi}^3}\psi+C\psi=0\, , \label{skdbv}
\end{equation}
where we set $\phi_1=\psi$ for simplicity. This Schamel equation represents an evolution equation for the OPdEASWs in the  magnetized collisional superthermal plasma under consideration, where the nonlinear coefficient $A$, which is responsible for the wave steepening, the dispersion coefficient $B$, which is responsible for the wave broadening, and the damping term $C$, which is associated with dissipation, are given by
\begin{eqnarray}
&&\hspace*{-0.5cm}A =-\frac{3}{4}\,\frac{\mathcal{S}_2\,v_p^3}{\beta\,l_z^2}\,,\label{non}\\
&&\hspace*{-0.5cm}B = \frac{v_p^3}{2l_z^2}\,\left[\frac{1}{\beta}+\frac{(1-l_z^2)}{\Omega_c^2}\right]\,,  \label{dis}\\
&&\hspace*{-0.5cm}C = \frac{\nu}{2}.  \label{dam}
\end{eqnarray}

\begin{figure}[!h]
\centering
\includegraphics[width=8.25cm]{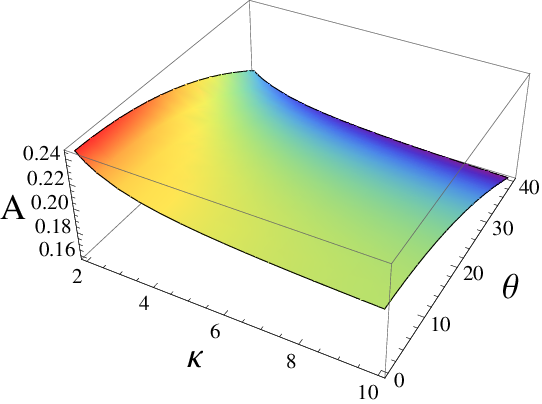}

\large{ (a)}
\vspace{0.5cm}

\includegraphics[width=8.25cm]{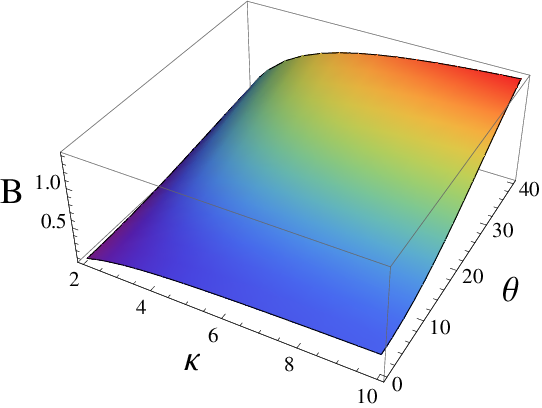}

\large{ (b)}
\caption{The nonlinear term $A$ (upper) and dispersion term $B$ (lower) versus $\kappa$ (a) where $\alpha=0.5$ and $\beta=0.5$ for different $\theta$, and (b) where $\beta=0.5$ and $\Omega_c=0.2.$} \label{Fig2}
\end{figure}
\begin{figure}[!h]
\centering
\includegraphics[width=8.25cm]{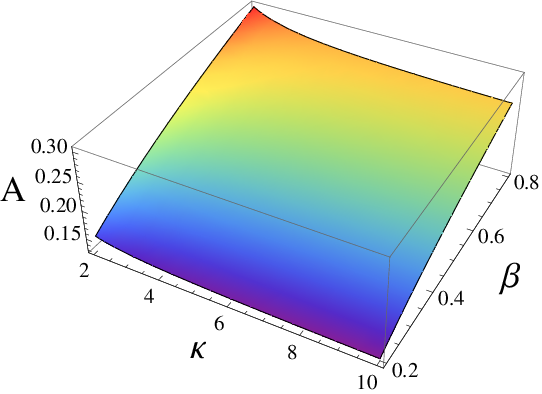}

\large{ (a)}
\vspace{0.5cm}

\includegraphics[width=8.25cm]{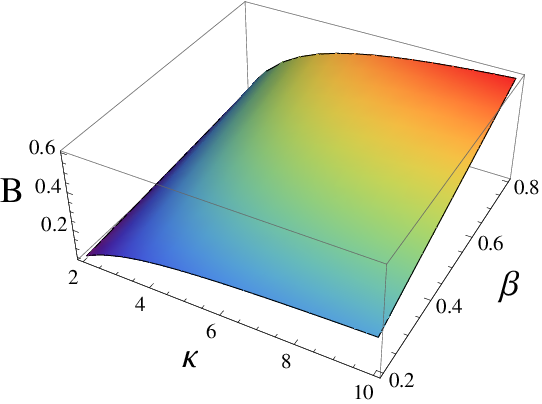}

\large{ (b)}
\caption{The nonlinear term $A$ (upper) and dispersion term $B$ (lower) versus $\kappa$ (a) where $\theta=10^{\circ}$ and $\alpha=0.5$ for different $\beta$, and (b) where $\theta=10^{\circ}$ and $\Omega_c=0.2$ for different $\beta$.} \label{Fig2a}
\end{figure}
When the nonlinear term $A$ balances the dispersive term $B$, the formation of OPdEASWs is possible, and thus the basic features of OPdEASWs are analysed depending on the parametric dependence of these two terms $A$ and $B$ of the considered plasma system. It is seen from equations (\ref{non}), (\ref{dis}), and (\ref{dam}) that the nonlinear coefficient $A$  (i.e., wave steepening) depends on $\kappa$, $\theta$, $\alpha$ and $\beta$, on the other hand, the dispersion coefficient $B$ (i.e., wave broadening) is an explicit function of $\kappa$, $\theta$, $\beta$, and $B_0$ as well as damping term $C$ is associated with $\nu$. That is, the wave steepening and broadening both are influenced by $\kappa$, $\theta$, and $\beta$, while $\alpha$ has a significant influence only on $A$ and only the wave broadening (via $B$) is affected by $B_0$. The propagation characteristics of EASWs in a magnetized collisional superthermal plasma with trapped electrons are far different than those in an unmagnetized collisionless superthermal plasma.
We have depicted the variation of the nonlinear coefficient $A$ and dispersion coefficient $B$ with $\kappa$ for different values of obliquity angle $\theta$  in Figure \ref{Fig2} and for different values of cold-to-hot electron number density at equilibrium $\beta$ in Figure \ref{Fig2a}.

\subsection{Solitary wave solution for OPdEASW\lowercase{s}}\label{easoln}
To identify the characteristics of the OPdEASWs, we now find the solitary wave solution of equation (\ref{skdbv}) for the magnetized collisional plasma by adopting the well-known hyperbolic tangent approach \cite{malfliet96}. In the absence of collisional damping (i.e., $\nu=0$), one may recover the usual KdV equation for EASWs in a magnetized plasma system. We assume a new travelling coordinate $\zeta=(\xi-u_0\tau)$, where $u_0$ is the wave speed in the reference frame (normalized by the hot electron sound speed $c_0$) and imposing the appropriate boundary conditions for the localized  solitary structures, namely $\psi\rightarrow 0$, $d_\zeta\psi\rightarrow 0$, and $d_{\zeta\zeta}\psi\rightarrow 0$ as $\zeta\rightarrow\pm \infty$,  we obtain the steady state solution of (\ref{skdbv}) in the following form
\begin{equation}
\psi(\xi,\tau)=\psi_0\,\rm{sech}^4\left[\frac{\zeta}{\delta}\right], \label{sol1}
\end{equation}
where $\psi_0=(3u_0/A)$ and $\delta=\sqrt{4B/u_0}$ are height and width of OPdEASWs, respectively.  We recall that $u_0$ is the incremental speed 
of OPdEASWs in the reference frame. Therefore, the actual speed of OPdEASWs in the plasma medium under consideration is $u_0+c_0$, i.e., the solitary wave is super-acoustic by nature. The solitary solution given in equation (\ref{sol1}) will be used as an initial condition to analyze the dissipative effect on EASWS.

Now, by assuming the presence of weak dissipation due to the electron-neutral collision, describes via the Schamel equation (\ref{skdbv}), one may find the time dependent electric potential electron acoustic solitary excitations in the form \cite{Ghosh2014,Karpman1977,Herman1990}
\begin{equation}
\psi(\xi,\tau)=\psi_m(\tau)\,\rm{sech}^4\sqrt{\frac{A\psi_m(\tau)}{12B}}\left[\xi-\frac{A}{3} \int_{0}^{\tau}\psi_m(\tau)d\tau\right]. \label{sol2}
\end{equation}

The above equation represents the electric potential excitations of the OPdEASWs, where the expressions for the time dependent height $\psi_m(\tau)$, width $\delta(\tau)$, speed $u(\tau)$ are, respectively

\begin{eqnarray} \left.
\begin{array}{l}
\psi_m(\tau)=\psi_0 e^{-2\nu\tau/3}, \\\\
\delta(\tau)=\sqrt{\frac{12B}{A\psi_0}}e^{\nu\tau/3}, \\\\
u(\tau)=\frac{A\psi_0}{3} e^{-2\nu\tau/3} , 
\end{array}
\right\} \label{ex}
\end{eqnarray}
where $\psi_0$ is the initial pulse amplitude at $\tau=0$, and the dissipative term $\nu$ measures the damping of the obliquely propagating solitary waves with time $\tau$.

It is clear in equations (\ref{non}) and (\ref{dis}) and also seen in Figures \ref{Fig2} that the nonlinear term $A$ (dispersion term $B$)  increase (decrease) with the obliqueness (via $\theta$). However, both $A$ and $B$ go up with the cold-to-hot electron number density at equilibrium $\beta$. That is, both the amplitude and width of OPdEASWs increase with the increasing values of $\theta$, but the width approaches zero and the amplitude becomes infinite when $\theta\rightarrow90^{\circ}$ for our considered plasma model suggesting that small obliquity angle $\theta$ need to be considered for numerical analysis since the assumption of large $\theta$ for which the waves are electrostatic is no longer valid, and one should then look for electromagnetic waves instead of electrostatic ones. It is also noted that for too strong obliqueness (i.e., for large $\theta$) the effect of $\mathbf{E\times B_0}$ (where $\mathbf{E}$ is the longitudinal wave electric field) can not be neglected and the electrostatic character of solitary excitations is violated \cite{verheest09}. We, therefore, consider a small oblique angle, where $0<\theta<45^\circ$ is considered for our numerical purposes in the manuscript.

\begin{figure}[!h]
\centering
\includegraphics[width=8.25cm]{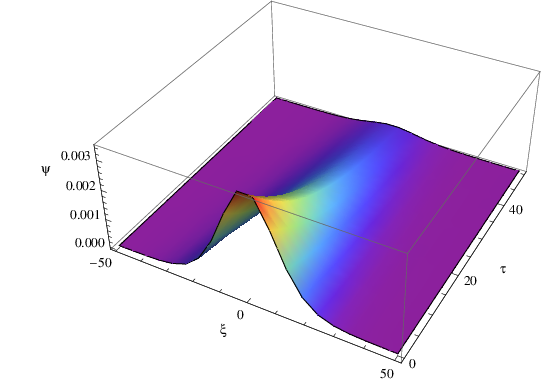}

\large{ (a)}
\vspace{0.5cm}

\includegraphics[width=8.25cm]{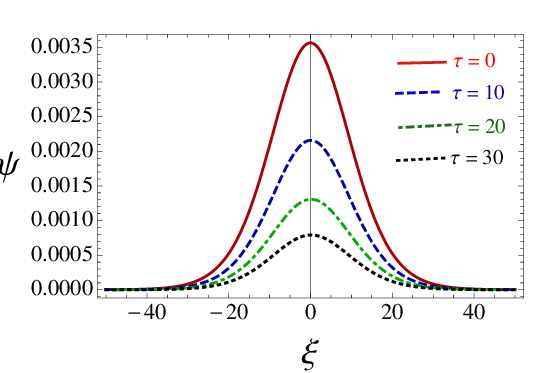}

\large{ (b)}\vspace{0.5cm}

\includegraphics[width=8.25cm]{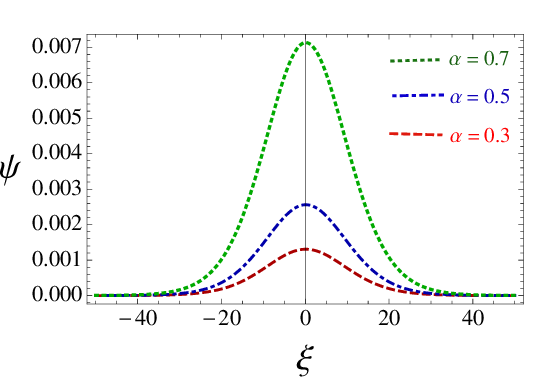}

\large{ (c)}
\caption{Showing (a) the nature of the electron-acoustic solitons in a magnetized collisional nonthermal ($\nu=0.1$) plasma for $\kappa=3,\,
\theta=10^{\circ},\,\alpha=0.3,\,\beta=0.5,\,\Omega_c=0.2,\,$ and $u_0=0.01$; (b) solitons profile for different time $\tau$; and (c) solitons profile for different $\alpha$ at $\tau=20.$}  \label{Fig3}
\end{figure}

\begin{figure}[!t]
\centering
\includegraphics[width=8.25cm]{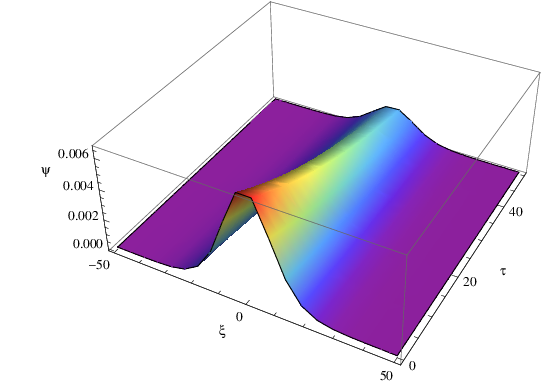}

\large{ (a)}
\vspace{0.5cm}

\includegraphics[width=8.25cm]{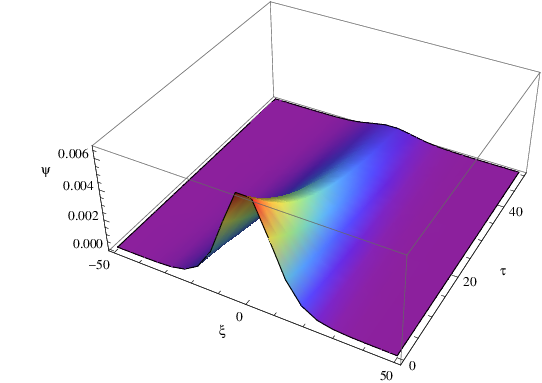}

\large{ (b)}


\caption{The nature of the dissipative electron-acoustic solitons in a magnetized collisional nonthermal  plasma for $\kappa=3,\,\theta=10^{\circ},\,\alpha=0.5,\,\beta=0.5,\,\Omega_c=0.5,\,u_0=0.01$, and (a) for $\nu=0.05$ and (b) for $\nu=0.1$.}

\label{Fig4}
\end{figure}


\begin{figure}[!h]
\centering
\includegraphics[width=8.25cm]{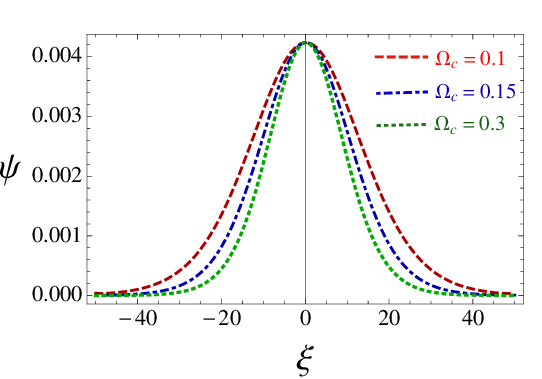}

\large{ (a)}
\vspace{0.5cm}

\includegraphics[width=8.25cm]{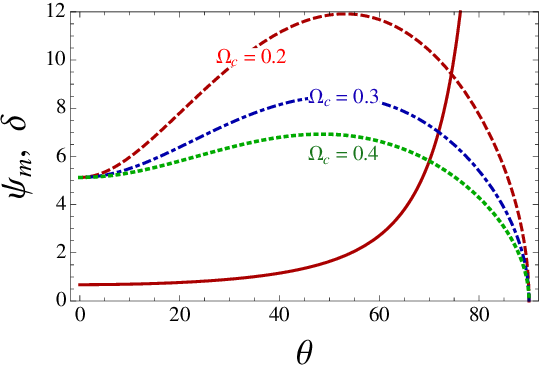}

\large{ (b)}
\caption{Showing (a) the dEASWs for different $\Omega_c$ for $\kappa=3,\, \theta=10^{\circ},\,\alpha=0.5,\,\beta=0.5,\,\tau=20,\,u_0=0.01$ and $\nu=0.05$; (b) the variation of amplitude $\psi_m$ and width $\delta$ of dEASWs with obliquity angle $\theta$.  Solid curve is for amplitude $\psi_m$ and others are for width $\delta$ for different $\Omega_c$. Other parameters are fixed at $\beta=0.5$, $\kappa=3$,  $\alpha=0.5$, $\nu=0.01$, and $u_0=0.1$.}
\label{Fig6}
\end{figure}


\subsection{Characteristics OPdEASW\lowercase{s}}\label{parainvest}
In this section, we are interested in the influence of different plasma configuration parameters e.g., the plasma superthermality, the concentration of trapped electron population, cold-to-hot electron number density, obliquity angle, external magnetic field, etc. on the dynamical properties of OPdEASWs in a collisional magnetized plasma system containing trapped superthermal electron populations. To identify the nature of OPdEASWs in terms of the different key plasma parameters, the solitary wave solution (\ref{sol1}) is considered as an initial condition to solve the Schamel equation (\ref{skdbv}) numerically via MATHEMATICA.

The nature of OPdEASWs with associated wave potentials in terms of time $\tau$ as well as trapping parameter $\alpha$ are depicted in Figure \ref{Fig3}. We show the effect of $\tau$ on OPdEASWs in Figures \ref{Fig3}(a,b), in which the amplitude (width) of the OPdEASWs decrease (constant) with the increase of time $\tau$, and the Figure \ref{Fig3}(c) indicates that the trapping is stronger for lower values of $\alpha$. A strongly superthermal (i.e., lower value of $\kappa$) magnetized collisional plasma is considered to examine the influence of $\alpha$, where other plasma parameters are fixed at $\tau=20$, $\theta=10^o$, $\beta=0.5$, $\Omega_c=0.2$, $u_0=0.01$, and $\nu=0.1$.  It is clear from equations (\ref{non}), (\ref{dis}), and Figure \ref{Fig3}(c) that the variation of $\alpha$ does have a significant influence on $A$ but does not have any effect on $B$, i.e., amplitude (width) of the OPdEASWs in a magnetized collisional superthermal is significantly influenced (remains same) by the variation of time (see in Figure \ref{Fig3}(b), and the amplitude (width) is found to rise (constant) with the increase of $\alpha$, as shown in Figure \ref{Fig3}(c).

We have analyzed the influence of electron-neutral collision via the collisional parameter $\nu$ on the nature of OPdEASWs. To do so, we have first considered the propagation of solitary solution (\ref{sol1}), the figure is omitted here (stable initial solitary pulse maintains its stability while propagating in a collisionless plasma medium with constant height and width), in a magnetized superthermal collisionless (i.e., $\nu=0$) plasma at an angle $10^{\circ}$ to that of the external magnetic field $B_0$ for $\kappa=3,\,\alpha=0.5,\,\beta=0.5,\,\Omega_c=0.5$, and $u_0=0.01$. The same initial pulse (\ref{sol1}) were then assumed to propagate in a magnetized superthermal collisional plasma (i.e., $\nu=0.05$) for the same plasma parametric values, which is depicted in Figure \ref{Fig4}(a). On the other hand, the initial solitary pulse becomes smaller (narrower) in amplitude (width) with time [see Figure \ref{Fig4}(b)] for $\nu=0.1$  and eventually diminishes as time passes.

Finally, we have examined the influence of obliquity angle $\theta$ and external magnetic field $B_0$ (via $\Omega_c$) on the basic features of OPdEASWs in Figure \ref{Fig6}. We recall here that both the nonlinear term $A$ (measures the steepness of waves) and dispersion term $B$ (measures the broadening of waves) is a function of $\theta$, while only $B$ depends on $\Omega_c$ -- suggesting the constant amplitude OPdEASWs for different $B_0$ (or $\Omega_c$). We have found that the external magnetic field does not have any effects on the amplitude of the OPdEASWs, but it does have significant effect on their which decreases with the increase in the magnitude of the external magnetic field (see Figure \ref{Fig6}), i. e., the external magnetic field makes the OPdEASWs spikier, which agrees with previous research work \cite{SSIK2012,mamun98,sultana19,sultana10} for different plasma contexts of space and laboratory plasmas.

\section{DISCUSSION} \label{Discussion}
In this manuscript, we have considered the oblique propagation of electrostatic dissipative electron-acoustic solitary waves in a three-component magnetized collisional plasma containing inertial cold electrons whose mass density provides the inertia, inertialess nonthermal hot electrons following $\kappa$ superthermal trapped distribution function whose thermal pressure provides the restoring force, and stationary ions. The electron-neutral  collisions, which is considered as the main source of dissipation, is taken into account. A nonlinear partial differential equation in the form of Schamel equation was derived, by adopting the well-known reductive perturbation technique, to model the evolution of the OPdEASWs in the plasma medium under consideration. The dispersive coefficient and the nonlinear coefficient were found to be positive for all plasma parameters, i.e., the considered plasma medium supports only positive potential OPdEASWs. We have analyzed the effects of different plasma configuration parameters [e.g., the superthermal effect by $\kappa$ (in Figure \ref{Fig1}), the obliqueness effect via $\theta$ in Figures \ref{Fig2} and \ref{Fig6}(b), effect of trapped particles via $\alpha$ depicted in Figure \ref{Fig3}(c), the dissipation effect via $\nu$ (see Figure \ref{Fig4}), magnetic field effect by $\Omega_c$ in Figure \ref{Fig6}, etc.)] on the dynamical properties of OPdEASWs. The results obtained in our theoretical investigation can be summarized as follows:

\begin{enumerate}
 \item{The basic properties such as amplitude, width, speed, stability, etc. of OPdEASWs  are influenced by the intrinsic plasma configuration parameters (e.g., plasma superthermality, the obliqueness, collisional parameter, and the external magnetic field).}

\item{The phase speed, $v_p$, is seen to higher for parallel propagating (i.e., $\theta=0^o$) EAWs than that for obliquely propagating EAWs (i.e., $0^o<\theta<45^o$), while $v_p$ is observed to higher in a more cold electron populated plasma than that in a less cold electron populated plasma.}

\item{The nonlinear term $A$ provides positive (i.e., $A>0$) for all plasma parametric values such as the magnetized collisional superthermal plasma system with two distinct temperature electrons under consideration may support only the propagation of positive potential electron-acoustic solitary structures.}

\item{The amplitude of OPdEASWs decreases in time, as expected, due to the effect of dissipation. This is true either for parallel propagation or oblique propagation. This is also true either for nonthermal (strongly superthermal) plasma or for thermal (Maxwellian or quasi-Maxwellian) plasma environments.}

\item{The time-dependent electron-acoustic solitary pulse amplitude is seen to larger, as expected, in a collision-free magnetized superthermal plasma than in a collisional magnetized superthermal plasma medium.}

\item{Larger in amplitude and wider in width electrostatic EASW is observed to form in a magnetized collisionless superthermal plasma than in a magnetized collisional Maxwellian plasma. On the other hand, the amplitude of OPdEASWs in a magnetized collisional superthermal plasma is found to increase with the increase in efficiency of trapped electron populations, but the width of OPdEASWs remains constant due to the variation of trapped electron populations.}

\item{The width of OPdEASWs is observed to increase (decrease) with the obliquity angle $\theta\leq45^{\circ}$ ($\theta\geq45^{\circ}$), and the amplitude of OPdEASw is seen to become zero as $\theta$ approaches $90^{\circ}$ -- suggesting the possibility for the propagation of electrostatic dEASWs for $0\leq\theta\leq45^{\circ}$.}

\item{The effect of the external magnetic field does not have any effect on the amplitude of the OPdEASWs, but it does have significant effect on their width, and their width decreases with the increase in the magnitude of the external magnetic field which agrees with existing published work \cite{SSIK2012,mamun98,sultana19,sultana10} in different contexts of space and laboratory plasmas.}


\end{enumerate}

By way of conclusion, we may say that our theoretical investigation should be useful for a better understanding of the characteristics of small but finite amplitude localized electrostatic disturbances that are ubiquitous in laboratory as well as space plasmas, where two distinct temperature electron populations are available, in presence of nonthermal electrons following trapped superthermal distribution and dissipation due to electron-neutral collision.

\section*{Acknowledgement}
M. R. Hassan is grateful to the Bangladesh Ministry of Science and Technology for awarding the National Science and Technology (NST) Fellowship.

\end{document}